\documentclass{aip-cp}

\usepackage[numbers]{natbib}
\usepackage{rotating}
\usepackage{graphicx}

\begin{document}

\title{Big Bang Nucleosynthesis, the CMB, and the Origin of Matter and Space-Time}

\author[aff1]{Grant J. Mathews\corref{cor1}}
\eaddress[url]{http://physics.nd.edu/people/faculty/grant-j-mathews/}
\author[aff1,aff2]{Mayukh Gangopadhyay}
\eaddress{mayukh.raj@saha.ac.in}
\author[aff1]{Nishanth Sasankan}
\eaddress{nsasanka@nd.edu}
\author[aff3]{Kiyotomo Ichiki}
\eaddress{ichiki.kiyotomo@c.mbox.nagoya-u.ac.jp}
\author[aff4,aff5,aff6]{Toshitaka Kajino}
\eaddress{kajino@nao.ac.jp}

\affil[aff1]{Center for Astrophysics, Department of Physics, University of Notre Dame, Notre Dame, IN 46556, USA}
\affil[aff2]{Theory Division, Saha Institute of Nuclear Physics, Bidhannagar, 700064, India}
\affil[aff3]{Department of Physics, Nagoya University, Nagoya 464-8602, Japan}
\affil[aff4]{National Astronomical Observatory, 2-21-1, Osawa, Mitaka, Tokyo 181-8588, Japan}
\affil[aff5]{Department of Astronomy, The University of Tokyo, 7-3-1 Hongo, Bunkyo-ku, Tokyo 113-0033, Japan}
\affil[aff6]{ International Research Center for Big-Bang Cosmology and Element Genesis, and School of Physics and Nuclear Energy Engineering, Beihang University, Beijing 100191, China}
\corresp[cor1]{Corresponding author: gmathews@nd.edu}

\maketitle

\begin{abstract}
 We summarize some applications of big bang nucleosythesis (BBN) and the cosmic microwave background (CMB) to constrain the first moments of the creation of matter in the universe.  We review the basic elements of BBN and how it constraints physics of the radiation-dominated epoch.  In particular, how the existence of higher dimensions  impacts the cosmic expansion through the projection of curvature from the higher dimension in the "dark radiation" term.  We summarize current constraints from BBN and the CMB on this brane-world dark radiation term. At the same time, the existence of extra dimensions during the earlier inflation impacts the tensor to scalar ratio and the running spectral index as measured in the CMB.  We summarize how the constraints on inflation shift when embedded in higher dimensions.  Finally, one expects that the universe was born out of a complicated multiverse landscape near the Planck time. In these moments the energy scale of superstrings was obtainable  during the early moments of chaotic inflation.  We summarize the quest for cosmological evidence of the birth of space-time out of the string theory landscape.   We will explore the possibility that a superstring excitations may have made itself known via a coupling to the field of inflation.  This may have left an imprint of "dips"  in the power spectrum of temperature fluctuations in the cosmic microwave background.  The identification of this particle as a superstring is possible because there may be evidence for different oscillator states of the same superstring that appear on different scales on the sky.  It will be shown that from this imprint one can deduce the mass, number of oscillations, and coupling constant for the superstring.  Although the evidence is marginal, this may constitute the first observation of a superstring in Nature.    
  \end{abstract}

\section{INTRODUCTION}
 We are at a unique period in the history of the human understanding of the cosmos. For the first time, we
have a clear picture of what the universe is comprised of, how long it has been in existence, and how it will evolve
in the future. This knowledge is the culmination of investigations via a number of cosmological probes including
supernovae, observations of the large scale distribution of galaxies and the inter-galactic medium, analysis of  the
cosmic microwave background, observations of the first stars of the early universe, and studies of the nucleosynthesis of the 
elements in the first few moments of cosmic expansion.  Here, we review some constraints  that big bang nucleosynthesis (BBN) and the Cosmic Microwave Background (CMB) 
place on models for the early universe.

\subsection{Physics of BBN}
If one adopts the notion that the universe is homogeneous and isotropic on a large enough scale, then space-time intervals are given by the Robertson-Walker metric,
\begin{equation}
ds^2 = -dt^2 + a(t)^2\biggl[\frac{dr^2}{1 - kr^2} + r^2d\theta^2 + r^2 \sin^2{\theta} d\phi^2\biggr]~~,
\end{equation}
where now $a(t)$ is the dimensionless scale factor and $k$ is the curvature parameter.  For this metric,  the evolution
of the early universe is simply given by the Friedmann equation which describes the the Hubble parameter H in terms
of densities $\rho$, curvature $k$, the cosmological constant $\Lambda$, and the cosmic scale factor $a$:
\begin{equation}
\biggl(\frac{\dot a}{a}\biggr)^2 \equiv H^2 = \frac{8}{3} \pi G \rho -\frac {k}{a^2} + \frac{\Lambda}{3}
 =  H_0^2\biggl[ \frac{\Omega_\gamma}{a^4}
 +\frac{\Omega_m}{a^3}
+ \frac{\Omega_k}{a^2}
+ \Omega_\Lambda \biggr]~~,
\label{Friedmann}
\end{equation}
where $H_0 = 67.74 \pm 0.46$ km sec$^{-1}$ Mpc$^{-1}$is the present value \cite{PlanckXIII}  of the Hubble parameter.  One can then define the various closure contributions
from relativistic particles, nonrelativistic matter, curvature, and dark energy:
\begin{equation}
\Omega_\gamma = 8 \pi G \rho_\gamma/(3 H_0^2)~~,~~ 
 \Omega_m = 8 \pi G \rho_m/(3 H_0^2)~~, ~~
 \Omega_k = k/(a^2 H_0^2)~,~~
  \Omega_\Lambda = \Lambda/(3 H_0^2)~~.
  \end{equation}

Based upon the {\it Planck} analysis \cite{PlanckXIII} then one can deduce that the contribution to closure from baryonic matter is  $\Omega_b  = 0.0486 \pm 0.0011$.  
The total matter content, however is $\Omega_m = 0.3089 \pm 0.0062$.  This
implies that a much larger fraction of the
universe is made of a completely unknown component of "cold dark matter," 
$\Omega_c = 0.260\pm 0.006$. Even more surprising
is that the universe is predominantly made of a completely exotic form of mass energy, i.e.~dark energy, denoted  as, 
$\Omega_\Lambda = 0.691 \pm 0.006$ for the case of a cosmological constant. 
In addition to these,  there is presently an almost negligible mass-energy contribution from relativistic
photons and neutrinos which we designate as 
$\Omega_\gamma  = \rho_\gamma/\rho_c =  5.46(19) \times 10^{-5.}$ The early universe includes the Planck epoch, the birth of space-time, inflation, reheating, a variety of cosmic
phase transitions (e.g. supersymmetry breaking, baryogenesis, the electroweak transition, and the QCD transition),
the epoch of big bang nucleosynthesis (BBN), and the production of the cosmic microwave background (CMB).
For most of the big bang only the radiation [$\Omega_\gamma$ term in Eq.~(\ref{Friedmann})] is important. There are, however, interesting
variants of big bang cosmology where this is not the case.

The precision with which these
parameters are now known is much better than a decade ago, however, one must keep in mind that these parameters are based upon the simplest
possible $\Lambda$CDM cosmology, and the analysis must be redone for more complicated cosmologies such as those described
below.

The only direct probe of the radiation dominated epoch
is the yield of light elements from BBN in the temperature regime from $10^8$ to $10^{10}$ K and times of about 1 to $10^3$
sec into the big bang. The only other probe is the spectrum of temperature 
fluctuations in the CMB which contains information of the first quantum 
fluctuations in the universe, and the details of the distribution and evolution of dark matter, baryonic
matter, photons and electrons near the time of the surface of photon last scattering (about $2.8 \times 10^5$ yr into the big bang).

One of the most powerful aspects of BBN is the simplicity\cite{Wagoner73,Yang84,Malaney93,Iocco09,Cyburt16,Mathews17} of the equations.  For the most part one can assume thermodynamic equilibrium of all species present.  The only non-equiulibrium processes are the thermonuclear reactions themselves.  
For all times relevant to standard  big-bang nucleosynthesis, one can ignore curvature and dark energy. Hence, the Friedmann equation is just:
\begin{equation}
\biggl(\frac{\dot a}{a}\biggr)=\sqrt{ \frac{8}{3} \pi G \rho_\gamma}  = H_0 \frac{\sqrt{\Omega_\gamma}}{a^2} \approx 1.13 T_{\rm MeV}^2 ~ {\rm sec}^{-1}  ~~.
\end{equation}
This leads to a simple relation for the evolution of the scale factor with time, $
a = \Omega_\gamma^{1/4} (2 H_0 t)^{1/2}$.
At the time of BBN the timescale for Compton scattering is short.  Hence, the number density of particles is simply given by Fermi-Dirac or Bose-Einstein distributions.

 In the standard big bang, with all chemical potentials $\mu_i$  set to zero, the total mass-energy density of the universe at the epoch of nucleosynthesis is given by
\begin{equation}
\rho = \rho_\gamma + \rho_{\nu_i} + \rho_i ~~,
\end{equation}
where $\rho_\gamma$, $\rho_{\nu_i}$, and $\rho_i$ 
are the energy densities due to photons, neutrinos, and charged leptons, respectively (including antiparticles).   The Friedmann equation  plus the fact that temperature decreases inversely with the scale factor, can be used to derive the relation between temperature and time during BBN.
\begin{equation}
 T \approx 1.4 g_{\rm eff}^{-1/4} \biggl( \frac{t}{\rm 1~sec}\biggr)^{-1/2}~{\rm MeV}~,
 \end{equation}
 where $g_{\rm eff}(T)$ is the effective number of relativistic degrees of freedom in bosons and fermions,
 \begin{equation}
 g_{\rm eff}(T) = \sum_{\rm Bose} g_{\rm Bose} + \frac{7}{8}\sum_{\rm Fermi} g_{\rm Fermi}  ~~.
\label{geff} 
\end{equation}

 The nuclear reactions, however, must be followed in detail as they fall out of equilibrium.  
 For nuclide $i$ undergoing reactions of the type 
$i + j \leftrightarrow j + k$ one can write:
 \begin{equation}
 \frac{dY_i}{dt} = \sum_{i,j,k} N_i \biggr( \frac{Y_l^{N_l} Y_k^{N_k}}{N_l ! N_k ! } n_k \langle \sigma_{l k} v \rangle - 
  \frac{Y_i^{N_i} Y_j^{N_j}}{N_i ! N_j ! } n_j \langle \sigma_{i j} v \rangle\biggr)
\label{nucrates}
 \end{equation}
where $Y_i = X_i/A_i$ is the abundance fraction,  $N_i$ is the number of reacting identical particles, $n_i$ is the number density of nucleus $i$ and $\langle  \sigma_{i j} v \rangle $
denoted the maxwellian averaged reaction cross section.
Once the forward reaction rate is known, the reverse reaction rate can be deduced from an application (cf. \cite{Mathews11}) of the principle of detailed balance. 

The reaction rates relevant to BBN have been  conveniently tabularized in analytic form in several sources \cite{Cyburt10,CF88,NACRE} and are a crucial ingredient to BBN calculations.  
   In all there are only
16 reactions of significance during BBN. \cite{Iocco09,Cyburt16,Coc17,Nakamura17,Foley17} See however \cite{Coc17} for suggestions on new reaction rates that may influence BBN.  Also, in Nakamura et al. \cite{Nakamura17} the broad range of reactions that can enter into the inhomogeneous big bang is summarized.  In order to be useful as a cosmological constraint one must know relevant nuclear reaction rates to very high precision ($\sim 1$\%).  Fortunately, unlike in stars, the energies at which these reactions occur in the early universe are directly accessible in laboratory experiments. 
Although considerable progress has been made \cite{Cyburt16,Coc17,Nakamura17,Foley17,Descouvemont04} toward quantifying and  reducing uncertainties in the relevant rates, much better rates are still needed for the neutron lifetime \cite{Serebrov10,Mathews05}, the $^2$H$(p,\gamma)^3$He, $^2$H$(d,n)^3$He, $^3$He$(d,p)^3$He, 
$^3$He$(\alpha,\gamma)^7$Be, and $^7$Be$(n,\alpha)^4$He reactions.

\section{Light Element Abundances}
One of the powers of standard-homogeneous BBN is that once the reaction rates are known, all of the light element abundances are determined
in terms of the single parameter $\eta$ (or $\eta_{10} $ defined as  the baryon-to-photon ratio in units of $10^{10}$). The crucial test of the
standard BBN is, therefore, whether the independently determined value of  $\eta_{10}$ from fits to the CMB reproduces all of the observed primordial
abundances. There are many reviews of this \cite{Iocco09,Cyburt16,Coc17,Nakamura17}, and also new constraints on the primordial
helium abundance \cite{Aver10,Aver15}. The best current abundance constraints can be adopted from Ref.~\cite{Cyburt16,Coc17,Nakamura17} as follows:

Deuterium is best measured in the spectra of narrow-line Lyman-$\alpha$  absorption systems in the foreground of high
redshift QSOs. Moreover, if one restricts the data to the six well resolved systems for which there are multiple 
Lyman-$\alpha$ lines\cite{Cyburt16,Coc17,Pettini12,Cooke14,Cooke16}, one obstains the presently adopted value \cite{Cyburt16} of 
$
{\rm D/H} = 2.53 \pm 0.04 \times 10^{-5} .
$

The abundance of $^3$He is best measured\cite{Bania02} in Galactic HII regions by the 8.665 GHz hyperfine transition of $^3$He$^+$. A
plateau with a relatively large dispersion with respect to metallicity has been found at a level of $^3$He/H $=(1.90 \pm 0.6) 
\times 10^{-5}$. However, it is not yet understood whether $^3$He has increased or decreased through the course of stellar and
galactic evolution \cite{Chiappini02,Vangioni-Flam03}. Fortunately, one can
avoid the ambiguity in galactic $^3$He production by making use of the fact that the sum of (D +$^3$He)/H is largely
unaffected by stellar processing. This leads to a reasonable 2$\sigma$ upper limit \cite{Iocco09,Coc17} of $^3$He/H $ < 1.7 \times 10^{-5}$.

The primordial $^4$He abundance, Yp is best determined from HII regions in metal poor irregular galaxies extrapolated
to zero metallicity.
In \cite{Aver10,Aver15} it was demonstrated that updated emissivities and the neutral hydrogen corrections generally increase the inferred
abundance, while the correlated uncertainties increase the uncertainty in the final extracted helium abundance.
This leads to:
$
{\rm Y_p }= 0.2449 \pm 0.0040 . 
$

\begin{figure}[t]
\label{fig:abund}
\centering
\includegraphics[scale=.7]{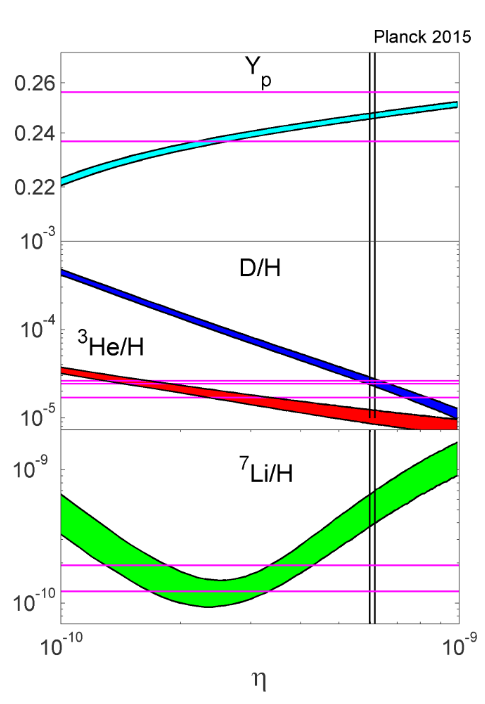}
\caption{BBN abundances as a function of the  baryon to photon ratio.  Shaded bands  correspond to the $2 \sigma$ (95\% C.L.) uncertainties deduced in 
[14].  
Horizontal lines show range of the uncertainties in the primordial abundances [6]. 
Vertical lines indicate the value of $\eta$ from the {\it Planck} analysis [1].
}
\end{figure}

The primordial abundance of $^7$Li is best determined from old metal-poor halo stars at temperatures corresponding
to the Spite plateau (see Refs.~[\cite{Iocco09,Coc17,Cyburt16}] and Refs. therein). There is, however, an uncertainty in this determination due to the
fact that the surface lithium in these stars may have experienced gradual depletion over the stellar lifetime via mixing with the higher temperature
 interiors. The best current limits are from Ref.~\cite{Cyburt16}.
$
 ^7{\rm Li/H} = 1.6 \pm 0.3 \times 10^{-10} ~~.
 $   

\section{Comparison of BBN with Observed abundances}

Ultimately, the value of BBN is in the detailed comparison between the  primordial abundances inferred from observation and the predictions based upon the BBN calculation.  Figure \ref{fig:abund} from \cite{Mathews17,Foley17} shows the current state of comparison.  The agreement between prediction and observation is quite good for most elements.
  There is, however, one glaring problem that remains.  The calculated and observed $^7$Li/H ratio differ by about a factor of 3.  This is known as "the lithium problem."  Many papers have addressed this problem, e.g.~\cite{Coc17,Nakamura17,Kusakabe17,Sato17,Yamazaki17}.  At present it is not yet known if this discrepancy derives from a destruction of Lithium on the old stars used to deduce the primordial Lithium abundance, or if it requires exotic new physics in the early universe \cite{Coc17,Nakamura17,Kusakabe17,Sato17,Yamazaki17}, or even a modification of the particle statistics in BBN itself \cite{He17}.  This disagreement may indicate new physics in the early universe.

\subsection{Is there Evidence for Large Extra Dimensions?}
In one form of the low-energy limit to M-theory \cite{Mtheory1,Mtheory2} the universe can be represented by two 10-dimensional manifolds separated by a large extra dimension. It is possible
that the extra dimension could manifest itself on the dynamics of the universe and BBN \cite{Ichiki02,Sasankan17}. For example, in a
Randall-Sundrum II \cite{Randall99} brane-world cosmology, the cosmic expansion for a 3-space embedded in a higher dimensional
space can be written \cite{Ichiki02,Sasankan17} as
\begin{equation}
\biggl( \frac{ \dot a}{a}\biggr)^2 = \frac{8}{3} \pi G \rho -\frac {k}{a^2} + \frac{\Lambda}{3} 
+ \frac{\kappa_5^4}{36} \rho^2 + \frac{\mu}{a^4}~~,
\label{Friedman2}
\end{equation}
where the four-dimensional gravitational constant $G_N$ is related to  the five-dimensional gravitational constant, $\kappa_5$, 
i.e. $G_N = {\kappa_5^4 \lambda}/{48 \pi}$, with $\lambda$ the intrinsic tension of the brane. The fourth term arises from the imposition of a junction
condition for the scale factor on the surface of the brane, and is not likely to be significant during BBN.  However it is significant during inflation \cite{Gangopadhyay17} as described below. 
The fifth  term, however,
scales just like radiation with a constant $\mu$ and is called the {\it dark radiation}. Hence, it can contribute durning BBN.  Its magnitude and sign derives from the
projection of curvature in higher dimensions onto four-dimensional space-time. Because this dark radiation scales as $a^{-4}$
it can affect both BBN and the CMB. It can significantly alter\cite{Ichiki02,Sasankan17} the fit to BBN abundances and the CMB, and hence can be constrained.
In \cite{Sasankan17} it was shown that the newest light element abundance constraints and reaction rates significantly limits the possibility of brane-world dark radiation.

\begin{figure}
\includegraphics[width=3.5 in]{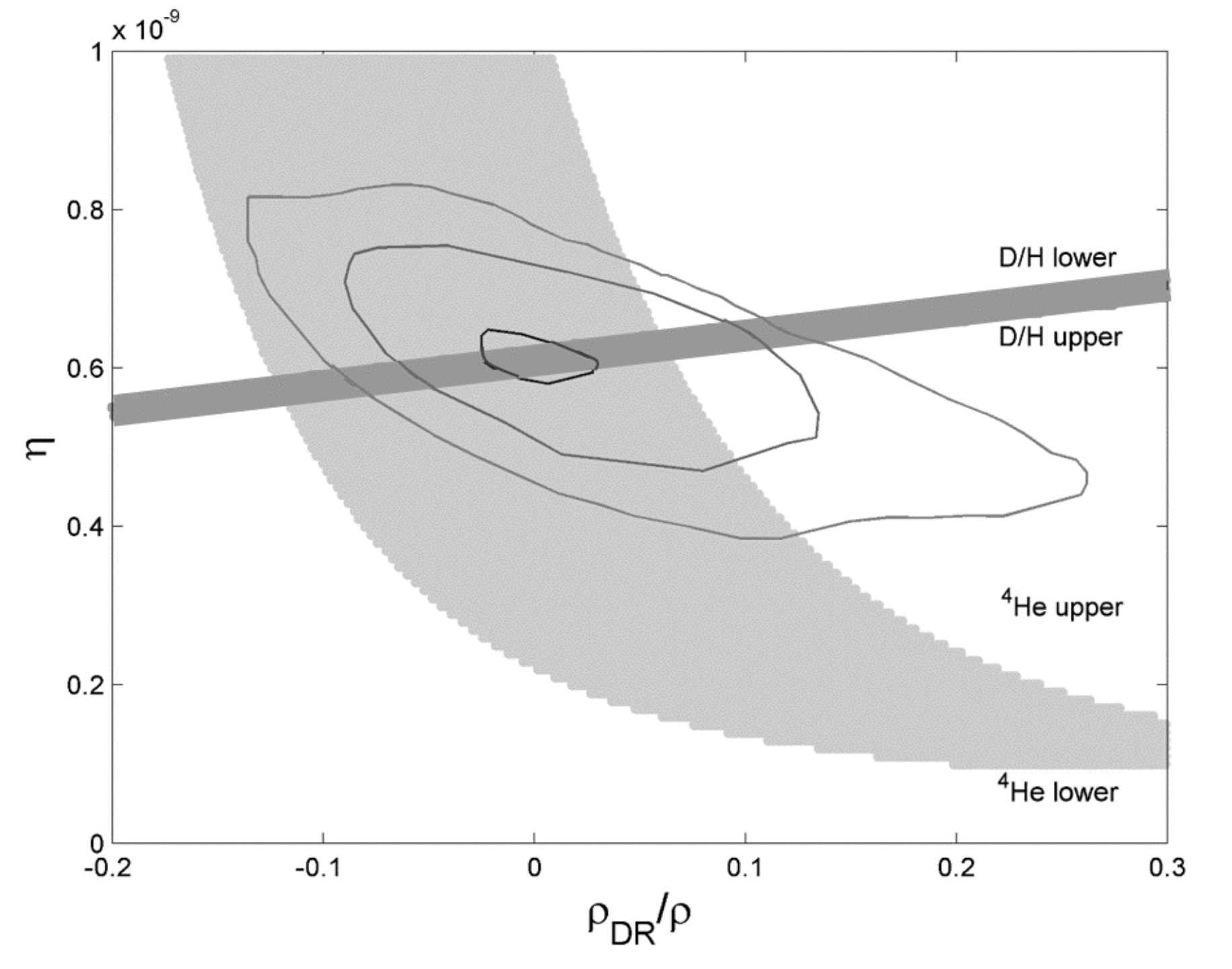} 
\caption{Constraints on  dark radiation in the $\rho_{\rm DR} /\rho$ (10 MeV) vs.~$\eta$  plane from Ref.~[33].  Contour lines  show the $1,~ 2, $  and $3 \sigma$ confidence limits based upon our fits to the CMB power spectrum. Dark shaded lines show the constraints from the primordial deuterium abundance as described in the text. The light shaded region shows the $Y_p$ constraint.}
\label{fig:4}
\end{figure}

 Figure \ref{fig:4} from \cite{Sasankan17} shows the combined constraints on $\eta$ vs.~dark radiation based upon our fits to both BBN and the CMB power spectrum. The  contour lines on Figure \ref{fig:4} show the CMB $1,~ 2,$  and $3 \sigma$ confidence limits in the $\eta$ vs.~dark radiation plane. The shaded regions show the BBN $Y_p$ and D/H constraints as labeled. 

The best fit concordance shown in Figure \ref{fig:4} is consistent with no dark radiation although in the BBN analysis there is  a slight  preference for  negative dark radiation.  A similar result was found in  the previous analysis of Ref.~\cite{Ichiki02}.  However, the magnitude of any dark radiation is much more constrained in the present analysis.  This can be traced to both the CMB and new light-element abundances.

\subsection{Constraints on Braneworld Inflation}
In \cite{Gangopadhyay17}  we have analyzed the $\rho^2$ term Randal Sundrum brane-world inflation scenario in the context of  the latest CMB constraints from {\it Planck}.  
The simplest explanation for the fact that the universe is so nearly 
flat today ($\Omega_{0} = 1.000 \pm 0.005$  Ref.~\cite{PlanckXIII}) and the near
isotropy of CMB is to conclude  that the universe has gone through an epoch of rapid inflation.\cite{Starobinsky,Guth, Linde}
 The simplest  view 
is that some vacuum energy $V(\phi)$ drives inflation due to the existence of a self-interacting scalar field $\phi$. That is, the
energy density of the cosmic  fluid in the early universe includes a dominant contribution from the evolution of a scalar field.
The mass energy density for a scalar field can be deduced from the Klein-Gordon Equation
\begin{equation}
\rho(\phi) = \frac{\dot \phi^2}{2} + \nabla^2 \phi   + V(\phi)~~.
\end{equation}
The inflaton  field $\phi$ itself evolves according to a damped harmonic-oscillator-like equation of motion:
\begin{equation}
\ddot \phi + 3 H \dot \phi - \nabla^2 \phi + dV/d\phi = 0~~.
\label{ddotphi}
\end{equation}
As the
universe expands, $H$ is large and the $\dot \phi$ term dominates as a kind of friction term. The universe is then temporarily   trapped in a slowly varying $V(\phi)$ dominated regime so that 
the scale factor grows exponentially.  This is known as the slow roll approximation for which in the absence of spatial fluctuations Eq.~(\ref{ddotphi}) becomes
\begin{equation}
\dot \phi =  \frac{dV/d\phi}{ 3 H}~~.
\end{equation}

The biggest unknown quantity in this paradigm is the form of $V(\phi)$. The simplest form $V(\phi) = (m/2) \phi^2$ may be motivated
by the Kahler potential in string theory, however, almost any form for the potential works well to describe the big bang. 
 Recently, however, the determination \cite{PlanckXX} of the ratio of the tensor to scalar contributions to the CMB power spectrum have ruled out many of the simplest forms of the 
 inflation generating potential.  Indeed, the only monomial type potentials that are marginally consistent with the data are those motivated by string theory
 such as the lowest order axion monodromy potentials\cite{McAllister10, Silverstein08} with $V(\phi) \propto \phi^{2/3}$ or $\propto \phi$, natural inflation, \cite{Freese90, Adams93, Freese93}
 or the $R^2$ inflation.\cite{Starobinsky80}

\begin{figure}[htb]
\includegraphics[width=6in,clip]{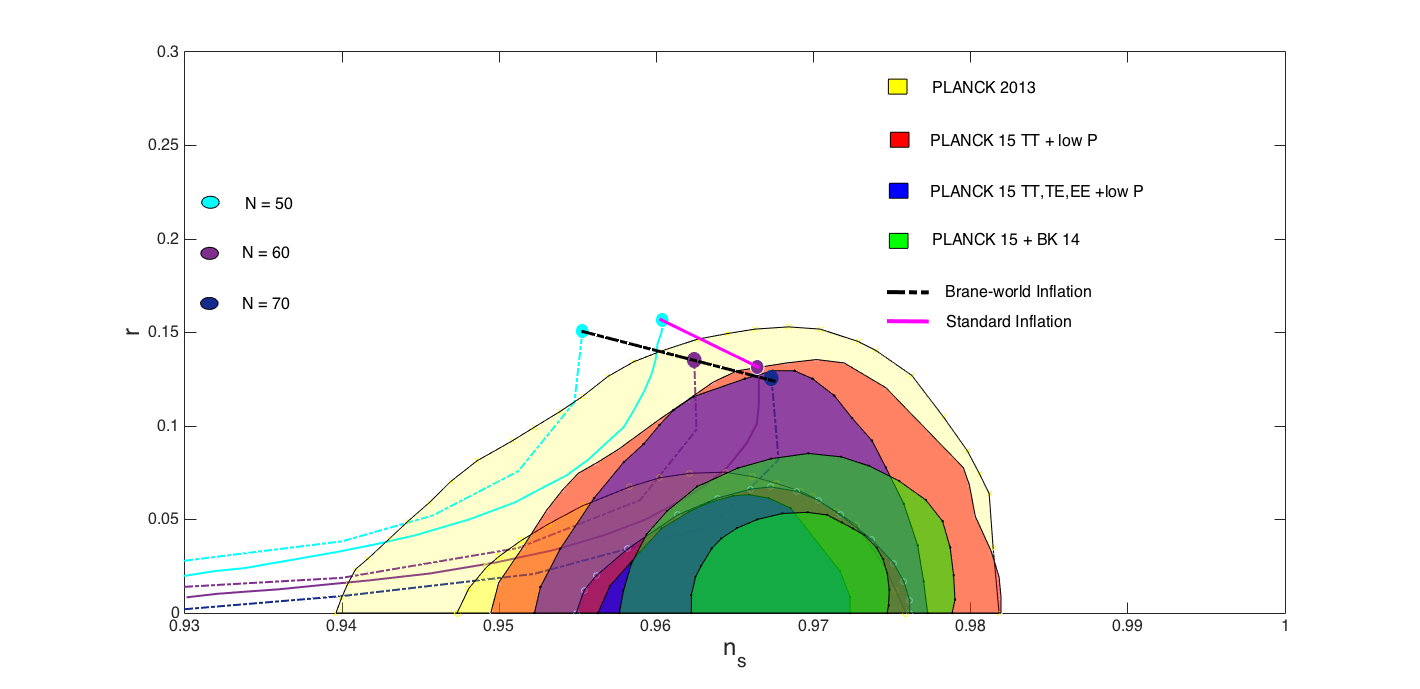} 
\caption{ Effects of brane-world cosmology in the case of natural inflation compared to  the standard inflation. The legend remains the same as in previous figures.  In this case, however,  the dashed lines denote the range of natural inflation solution.}
\label{fig:14}
\end{figure}

In \cite{Gangopadhyay17} we summarized constraints on the most popular classes of models and explored some more realistic inflaton effective potentials both in normal inflation and including the $\rho^2$ term in Eq.~[\ref{Friedman2}]. We could confirm that in general the brane-world scenario increases the tensor-to-scalar ratio, thus making this paradigm less consistent with the {\it Planck} constraints.  
 However, in the case of the lowest order $\phi^{2/3}$ axion monodromy there can be a slight improvement over the standard model in the case of a large number of $e$-folds $N$, particularly when compared to the {\it Planck 15}  TT + low-P constraints.
 Indeed, when BICEP2/Keck constraints are included, all monomial potentials in the brane-world scenario become disfavored compared to the standard scenario.  However,  for natural inflation the brane-world scenario fits the constraints better due to larger allowed values of $e$-foldings $N$ before the end of inflation in the brane-world.  Figure \ref{fig:14} from \cite{Gangopadhyay17} shows the tensor-to scalar ratio vs. the spectral index for inflation fluctuations for the case of natural inflation.  The fact that the inflation is more rapid in the brane-world allows for more e-folds of inflation and a better fit to the CMB constraints.
 
\section{Are there observable effects of string excitations during inflation?}
In \cite{Mathews15,Mathews17b} we have explored the possibility  that both the suppression of the $\ell = 2$ multipole moment of the power spectrum of the cosmic microwave background temperature fluctuations and the possible dip in the power spectrum for $\ell = 10-30$  as well as a possible dip for $\ell \approx 60$ could be explained.
 These features could arise from the resonant creation of  sequential excitations of a  fermionic (or bosonic)  closed superstring that couples
 to the inflaton field.  We utilize the example of  a D=26 closed bosonic string with one toroidal compact dimension as an illustration of how string excitations might imprint themselves on the power spectrum of the CMB.  We considered the possibilities of successive momentum states, winding states, or oscillations on the string as the source of the three possible dips in the power spectrum.  It was concluded that sequential momentum states may provide the best fit for these three resonances.
Although the evidence of the dips at $\ell \approx 20 $ and $\ell \approx 60$ are of
marginal statistical significance,  and there are other possible interpretations of these features, this could constitute  the first  observational evidence of the existence of a superstring in Nature.

\begin{figure}[htb]
\includegraphics[width=3.5in,clip]{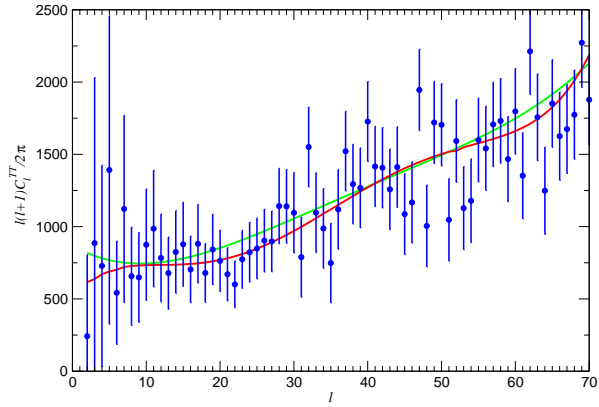} 
\caption{ The fit (red line) to $\ell \approx 2$, $\ell \approx 20$ and  $\ell \approx 60$ suppression of the {\it TT}  CMB power spectrum based upon a superstring coupled to the inflaton \cite{Mathews17b}. Points with error bars are from the  {\it Planck} Data Release \cite{PlanckXIII}.  The green line shows the best standard $\Lambda$CDM power-law fit to the {\it Planck} CMB power spectrum}
\label{fig:1}
\end{figure}

Figure \ref{fig:1} illustrates  the best  fit to the {\it TT} CMB power spectrum that includes both the ${\ell } \approx  2$,  ${\ell } \approx  20$ and ${\ell } \approx  60$ suppression of the CMB.

In that analysis, we  deduced the following resonance parameters:
$$\ell \approx 2, ~A = 1.7\pm 1.5, ~k_* (n+1)= 0.0004 \pm 0.0003 ~h~{\rm Mpc}^{-1}$$
$$\ell \approx 20,~A = 1.7\pm 1.5 ,~k_*(n) = 0.0015 \pm 0.0005  ~h~{\rm Mpc}^{-1}$$
$$\ell \approx  60,~A = 1.7 \pm 1.5,~k_*(n-1) = 0.005 \pm 0.0004  ~h~{\rm Mpc}^{-1~~,}$$
where $A$ is the amplitude of the dip in the primordial power spectrum and $k_*$ is the wave number associated with the resonant particle production.
From the relations between in the mass of the string excitation and the wave number, we deduce a mass of $m \sim  (8-11)  ~\frac{m_{\rm pl}}{\lambda^{3/2}}$ for the state at $\ell \approx 20$.
From the ratios of masses among the three resonances, we conclude that these resonances are most consistent with successive momentum states (or winding states) on a superstring.

\section{ACKNOWLEDGMENTS}
Work at the University of Notre Dame is supported
by the U.S. Department of Energy under 
Nuclear Theory Grant DE-FG02-95-ER40934.


\nocite{*}
\bibliographystyle{aipnum-cp}%

\begin{thebibliography}{99}
\bibitem{PlanckXIII} {\it Planck XIII} Collaboration, Astron. \& Astrophys. {\bf 594} A13 (2016).
%
\bibitem{Wagoner73} R.V. Wagoner, Ap.J. 179 (1973) 343.
\bibitem{Yang84} J. Yang, M.S. Turner, G. Steigman, D.N. Schramm, and K.A. Olive, Ap.J. 281 493 (1984).
\bibitem{Malaney93} R. A. Malaney and G. J. Mathews, G. J., Phys. Rep. 229 , 145 (1993).
\bibitem{Iocco09} F. Iocco, et al., Phys. Rep., 472, 1 (2009). 
\bibitem{Cyburt16} R. H. Cyburt, B. D. Fields, and K. A. Olive, Rev. Mod. Phys., 88, 015004 (2016).
\bibitem{Mathews17} G.~J.~Mathews, M. Kusakabe, and T. Kajino, Int. J. Mod. Phys. E26, 1741001 (2017).
\bibitem{Mathews11} G. J. Mathews, Y. Pehlivan, T. Kajino, A. B. Balantekin and M. Kusakabe, " Astrophys. J., 727, 10 (2011).
\bibitem{Cyburt10} R.~H.~Cyburt, et al., Astrophys. J.~Suppl. {\bf 189}, 240 (2010).
\bibitem{CF88}Caughlan, G. R. \& Fowler, W. A. 1988, ADNDT, 40, 283
\bibitem{NACRE} Angulo,~C.~et al. 1999, Nucl. Phys. A., 656, 3 (NACRE)
\bibitem{Coc17}  A. Coc and E. Vangioni, Int. J. Mod. Phys. E 26 (2017) 1741002.
\bibitem{Nakamura17}R. Nakamura, M.-A. Hashimoto, R. Ichimasa and K. Arai, Int. J. Mod. Phys. E 26,  1741003 (2017).
\bibitem{Foley17} M. Foley, N. Sasankan, M. Kusakabe and G. J. Mathews, Int. J. Mod. Phys. E 26,  1741008 (2017).
\bibitem{Descouvemont04}P.~Descouvemont, et al., ADNDT, 88, 203 (2004).
\bibitem{Serebrov10} A. P. Serebrov and A. K. Fomin, Phys. Rev. C 82, 035501) (2010).
\bibitem{Mathews05} G. J. Mathews T. Kajino, and T. Shima, Phys. Rev. D71, 021302 (2005).
\bibitem{Aver10} E. Aver, K. A. Olive, and E. D. Skillman, JCAP 05 003, (2010), JCAP 11, 017 (2013).
\bibitem{Aver15} E. Aver, K. A. Olive, and E. D. Skillman, arXiv: 1503:08146 (2015).
\bibitem{Pettini12} M. Pettini \& Cooke, MNRAS, 425, 2477 (2012).
\bibitem{Cooke14} R. Cooke, et al. Astrophys. J., 781, 31 (2014)
\bibitem{Cooke16} R. Cooke, et al. Astrophys. J., {\it in press} (2016)
\bibitem{Bania02} T. M. Bania, R. T. Rood and D. S. Balser, Nature 415, 54 (2002).
\bibitem{Chiappini02}C. Chiappini, A. Renda and F. Matteucci, Astron. Astrophys. 395, 789 (2002).
\bibitem{Vangioni-Flam03}E. Vangioni-Flam, K. A. Olive, B. D. Fields and M. Casse, Astrophys. J. 585, 611 (2003).
\bibitem{Kusakabe17}  M. Kusakabe, G. J. Mathews, T. Kajino and M.-K. Cheoun, Int. J. Mod. Phys. E26,  1741004 (2017).
\bibitem{Sato17}  J. Sato, T. Shimomura and M. Yamanaka, Int. J. Mod. Phys. E 26,  1741005 (2017).
\bibitem{Yamazaki17}  D. Yamazaki, M. Kusakabe, T. Kajino, G. J. Mathews and M.-K. Cheoun, Int. J. Mod. Phys. E 26,  1741006 (2017).
\bibitem{He17} S. Q. Hou, J. J.  He, J. J. A. Parikh, D.  Kahl, D. C. A. Bertulani, T.  Kajino, G. J.  Mathews, and G.  Zhao,  Astrophys. J., 834, 165 (2017).
\bibitem{Mtheory1}  P. Horava, and E. Witten,  Nuclear Physics B. 460 (3): 506Ð524 (1996a). 
\bibitem{Mtheory2} P. Horava and E. Witten, Edward (1996b). Nuclear Physics B. 475 (1): 94Ð114 (1996b).
\bibitem{Ichiki02} K. Ichiki, M. Yahiro, T. Kajino, M. Orito, G. J. Mathews, Phys. Rev. D66, 043521 (2002).
\bibitem{Sasankan17} N. Sasankan, M. Gangopadhyay, M. Kusakabe and G. G. Mathews, Phys. Rev. D95, 083516 (2017).
\bibitem{Randall99} L. Randall and R. Sundrum, Phys. Rev. Lett. 83, 3370 (1999); 83, 4690 (1999).
\bibitem{Gangopadhyay17} M. Gangopadhyay and  G. J. Mathews, J. Cosmology and Astrophysics (JACP)  (2017)  arXiv:161105123.
\bibitem{PlanckXX} {\it Planck XX} Collaboration, Astron. \& Astrophys. {\bf 594} (2015) A20 (2016).
\bibitem{Starobinsky} A. Starobinsky,  Soviet Physics JETPL, 30, 682 (1979).
\bibitem{Guth} A.  Guth,  PRD, 23, 347 (1981).
\bibitem{Linde} A. Linde,  PLB, 108, 393 (1982).
\bibitem{McAllister10} L. McAllister, E. Silverstein and  A.  Westphal, Phys.Rev., {\bf D82}, 046003 (2010).
\bibitem{Silverstein08}E. Silverstein, and A.  Westphal,  Phys.Rev., {\bf D78}, 106003 (2008).
\bibitem{Freese90} K. Freese, J. A. Frieman and A. V. Olinto, Phys. Rev. Lett. {\bf 65}  3233 (1990).
\bibitem{Adams93} F. C. Adams, J. R. Bond, K. Freese, J. A. Frieman and A. V. Olinto, Phys. Rev. {\bf D47}, 426  (1993).
\bibitem{Freese93} K. Freese and W. H. Kinney, Phys. Rev. {\bf D70}, 083512 (2004).
\bibitem{Starobinsky80} A. Starobinsky,  Phys. Lett., B91, 99 (1980). 
\bibitem{Mathews15} G.~J.~Mathews, M. R. Gangopadhyay, K. Ichiki, and T. Kajino, Phys. Rev. D92, 123519 (2015).
\bibitem{Mathews17b} G.~J.~Mathews, M. R. Gangopadhyay, K. Ichiki, and T. Kajino, Phys. Rev. D Submitted (2017).

 \end{thebibliography}

\end{document}